# A Hop-by-Hop Congestion-Aware Routing Protocol for Heterogeneous Mobile Ad-hoc Networks

S.Santhosh baboo
PG & Research Department of computer application
DG Vaishnav College, Arumbakkam
Chennai, India

B.Narasimhan
Department of BCA
K.G.College of Arts& Science, Saravanam Patti
Coimbatore-35, India
bnarasimhanphd@gmail.com

***Abstract*—In Heterogeneous mobile ad hoc networks (MANETs) congestion occurs with limited resources. Due to the shared wireless channel and dynamic topology, packet transmissions suffer from interference and fading. In heterogeneous ad hoc networks, throughput via a given route is depending on the minimum data rate of all its links. In a route of links with various data rates, if a high data rate node forwards more traffic to a low data rate node, there is a chance of congestion, which leads to long queuing delays in such routes. Since hop count is used as a routing metric in traditional routing, it do not adapt well to mobile nodes. A congestion-aware routing metric for MANETs should incorporate transmission capability, reliability, and congestion around a link. In this paper, we propose to develop a hop-by-hop congestion aware routing protocol which employs a combined weight value as a routing metric, based on the data rate, queuing delay, link quality and MAC overhead. Among the discovered routes, the route with minimum cost index is selected, which is based on the node weight of all the in-network nodes. Simulation results prove that our proposed routing protocol attains high throughput and packet delivery ratio, by reducing the packet drop and delay.***

## I. INTRODUCTION

### A. Heterogeneous Ad-hoc Wireless Networks

An ad hoc network is also called as infrastructure less networks which is a collection of mobile nodes which forms a temporary network without the help of central administration or standard support devices regularly available in conventional networks. Mobile ad hoc wireless networks have the ability to establish networks at anytime, anywhere to possess the assurance of the future. These networks do not depend on irrelevant hardware because it makes them ideal candidate for rescue and emergency operations. The constituent wireless nodes of these network build, operate and maintain these networks. Each node asks the help of its neighboring nodes to forward packets because these nodes usually have only a limited transmission range.

A homogeneous ad hoc network suffers from poor scalability because the network performance is degraded quickly as the number of nodes increases. The nodes are usually heterogeneous in realistic ad hoc networks. For example, in a battlefield network, portable wireless devices are carried by soldiers, and more powerful and reliable communication devices are carried by vehicles, tanks, aircraft, and satellites and these devices/nodes have different communication characteristics in terms of transmission power, data rate, processing capability, reliability, etc. Hence it would be more realistic to model these network elements as different types of nodes [1]. Such heterogeneous networks nodes are portable to transmit at different power levels and thus cause communication links of varying ranges.

### B. Routing in Mobile Ad-hoc Wireless Networks

Specially configured routing protocols are engaged in order to establish routes between nodes which are more than a single hop. The ability to trace routes in spite of a dynamic topology is the unique feature of these protocols. These protocols can be categorized into two main types: Reactive (On-demand) and Proactive (Table-driven). Evaluating the routes continuously within the network is done by proactive protocols, so when a packet needs to be forwarded the route is already known and can be immediately used. . Reactive protocols appeal to a route determination procedure on demand only.

### C. Congestion in Mobile Ad-hoc Wireless Networks

In mobile ad hoc networks (MANETs) congestion occurs with limited resources. Due to the shared wireless channel and dynamic topology, packet transmissions suffer from interference and fading, in such networks. The network load is burdened through the transmission errors. There is an increasing demand for support of multimedia communications in MANETs, recently. Large amount of real-time traffic involves high bandwidth and it is liable to congestion. Congestion leads to packet losses and bandwidth degradation and also wastes time and energy on congestion recovery.

## II. RELATED WORK

Xiaoqin Chen et al. [2] have proposed a congestion-aware routing metric which was employed data-rate, MAC overhead, and buffer queuing delay, with preference given to less congested high throughput links to improve channel utilization also they have proposed the Congestion Aware Routing protocol for Mobile ad hoc networks (CARM). CARM has applied a link data-rate categorization approach to prevent routes with mismatched link data-rates. CARM was only discussed and simulated in relation IEEE 802.11b networks; however, it was applied to any multi-rate ad hoc network.

Ming Yu et al. [3] have proposed a link availability-based QoS-aware (LABQ) routing protocol for mobile ad hoc networks based on mobility prediction and link quality measurement, in addition to energy consumption estimate. They have provided highly reliable and better communication links with energy-efficiency.

Yung Yi and Sanjay Shakkottai [4] have developed a fair hop-by-hop congestion control algorithm with the MAC constraint was being imposed in the form of a channel access time constraint, using an optimization-based framework. In the absence of delay, they have shown that this algorithm was globally stable using a Lyapunov-function-based approach and in the presence of delay, they have shown that the hop-by-hop control algorithm has the property of spatial spreading.

R.Asokan et al. [5] were being extended the scope to QoS routing procedure, to inform the source about QoS available to any destination in the wireless network. However, existing QoS routing solutions were dealt with only one or two of the QoS parameters. It was important that MANETs was provided QoS support routing, such as acceptable delay, jitter and energy in the case of multimedia and real time applications. They have proposed a QoS Dynamic Source Routing (DSR) protocol using Ant Colony Optimization (ACO) called Ant DSR (ADSR).

Lei Chen and Wendi B. Heinzelman [6] have proposed a QoS-aware routing protocol that were incorporated an admission control scheme and a feedback scheme to meet the QoS requirements of real-time applications. The novel part of this QoS-aware routing protocol was the use of the approximate bandwidth estimation to react to network traffic. They have implemented these schemes by using two bandwidth estimation methods to find the residual bandwidth available at each node to support new streams.

Chenxi Zhu and M. Scott Corson [7] have developed a QoS routing protocol for ad hoc networks using TDMA. Their object was to establish bandwidth guaranteed QoS routes in small networks whose topologies were changed at low to medium rate. The protocol was based on AODV, and built QoS routes only as needed. They have started with the problem of calculating the available bandwidth on a given route and develop an efficient algorithm and then they were used the algorithm in conjunction with AODV to perform QoS routing.

Duc A. Tran and Harish Raghavendra [8] have proposed CRP, a congestion-adaptive routing protocol for MANETs. CRP tried to prevent congestion from occurring in the first place, rather than dealing with it reactively. A key in CRP design was the bypass concept. A bypass was a sub path connecting a node and the next non congested node. If a node was aware of a potential congestion ahead, it was found a bypass that was used in case the congestion actually occurred or. Part of the incoming traffic was sent on the bypass, was being made the traffic was being come to the potentially congested node less. The congestion was avoided as a result.

RamaChandran and Shanmugavel [11] have proposed and studied three cross-layer designs among physical, medium access control and routing (network) layers, using Received Signal Strength (RSS) as cross-layer interaction parameter for energy conservation, unidirectional link rejection and reliable route formation in mobile ad hoc networks.

Jitendra Padhye et al. [12] have considered the problem of estimating pairwise interference among links in a multi-hop wireless testbed. Using experiments done in a 22-node, 802.11- based testbed, they have shown that some of the previously proposed heuristics for predicting pairwise interference were inaccurate. They have then proposed a simple, empirical methodology to estimate pairwise interference using only O (n2) measurements. They have shown that their methodology accurately predicted pairwise interference among links in their testbed in a variety of settings. Their methodology is applicable to any 802.11-based wireless network where nodes use omni-directional antennas.

Xinsheng Xia et al. [14] have introduced a method for cross-layer design in mobile ad hoc networks. They have used fuzzy logic system (FLS) to coordinate physical layer, datalink layer and application layer for cross-layer design. Ground speed, average delay and packets successful transmission ratio were selected as antecedents for the FLS. The output of FLS has provided adjusting factors for the AMC (Adaptive Modulation and Coding), transmission power, retransmission times and rate control decision.

Congestion in mobile ad hoc networks leads to transmission delays and packet loss, and causes wastage of time and energy on recovery. Routing protocols which are adaptive to the congestion status of a mobile ad hoc network can greatly improve the network performance. Xiaoqin Chen et al. [15] have proposed a congestion-aware routing protocol for mobile ad hoc networks which has used a metric incorporating data-rate, MAC overhead, and buffer delay to combat congestion. This metric was used, together with the avoidance of mismatched link data-rate routes, to make mobile ad hoc networks robust and adaptive to congestion.

Ming Yu et al. [16] have proposed a link availability-based QoS-aware (LABQ) routing protocol for mobile ad hoc networks based on mobility prediction and link quality measurement, in addition to energy consumption estimate. There goal was to provide highly reliable and better communication links with energy-efficiency. It has also reduced the average end-to-end delay for data transfer and it has enhanced the lifetime of nodes by making energy-efficient routing decisions.

## III. Proposed Work

### *A. Congestion Control in Mobile Adhoc Network*

Congestion in wireless networks is slightly different from that of wired networks. The following are the general cause of congestion:

1. The throughput of all nodes in a particular area gets reduced because many nodes within range of one another attempt to transmit simultaneously, resulting in losses.

2. The queue or buffer used to hold packets to be transmitted, overflows within a particular node. This is also the cause of losses.

3. Moreover in a heterogeneous network, different data-rates will almost certainly lead to some routes having different links with quite different data-rates. The packets will build up at the node heading the lower data-rate link which leads to long queuing delays. Link reliability is the additional cause for the congestion. Congestion is increased due to packet retransmission [2], if the link breaks.

Since hop count is used as a routing metric in traditional routing, it do not adapt well to mobile nodes. To counter node mobility, many existing routing schemes use message exchanges, such as hello packets. Unless a link is broken these schemes do not change the routes, rather than taking precautions and make sure the link would not be broken [3].

### *B. Protocol Overview*

A congestion-aware routing metric for MANETs should incorporate transmission capability, reliability, and congestion around a link.

In this paper, we propose to develop a Hop-by-hop congestion aware routing protocol which employs the following routing metrics:

- Data-rate
- Buffer queuing delay
- Link Quality
- MAC Overhead

With preference given to less congested high throughput links to improve channel utilization.

In our proposed routing protocol, after the estimating the above metrics, a combined weight value is calculated for each node. We select any multi path on-demand routing protocol, which discovers multiple disjoint routes from a source to destination, as our basis. Among the discovered routes, the route with minimum cost index is selected, which is based on the node weight of all the in-network nodes for each packet successfully delivered from the source node to the destination node. The node's cost index is calculated in a backward propagating way. The cost indices of a node's possible downstream neighbors are obtained by the feedbacks of its downstream neighbors.

### *C. Link Quality Estimation*

To be able to see that a node is moving and a route is about to break, we relay on the fact that communication is based on electronic signals. Because of that it is possible to measure the quality of the signal and based on that guess if the link is about to break. This can be used by the physical layer to indicate to the upper layer when a packet is received from a host, that is sending with a signal lower than a specific value and then indicate that that node is in pre-emptive zone [9],[10]. Thus, using the received signal strength from physical layer, link quality can be predicted and links with low signal strength will be discarded from the route selection.

When a sending node broadcasting RTS packet, it piggybacks its transmissions power $P_t$. On receiving the RTS packet, the intended node measures the signal strength received which holds the following relationship for free-space propagation model [11].

$$P_r = P_t.(\lambda / 4\pi d)^2.G_t.G_r \qquad (1)$$

Where $\lambda$ is the wavelength carrier, $d$ is distance between sender and receiver, $G_t$ and $G_r$ are unity gain of transmitting and receiving omni directional antennas, respectively. The effects of noise and fading are not considered.

So, the link quality $L_q = P_r$ (2)

### *D. Estimating MAC Overhead*

In this network, we consider IEEE 802.11 MAC with the distributed coordination function (DCF). It has the packet sequence as request-to-send (RTS), clear-to-send (CTS), and data, acknowledge (ACK). The amount of time between the receipt of one packet and the transmission of the next is called a short inter frame space (SIFS). Then the channel occupation due to MAC contention will be

$$C_{occ} = t_{RTS} + t_{CTS} + 3t_{SIFS} \qquad (3)$$

Where $t_{RTS}$ and $t_{CTS}$ are the time consumed on RTS and CTS, respectively and $t_{SIFS}$ is the SIFS period.

Then the MAC overhead OHMAC can be represented as

$$OH_{MAC} = C_{occ} + t_{acc} \qquad (4)$$

Where $t_{acc}$ is the time taken due to access contention.

The amount of MAC overhead is mainly dependent upon the medium access contention, and the number of packet collisions. That is, $OH_{MAC}$ is strongly related to the congestion around a given node.

$OH_{MAC}$ can become relatively large if congestion is incurred and not controlled, and it can dramatically decrease the capacity of a congested link.

### *E. Estimating End to End Delay*

There is a significant variation between the end-to-end delay reported by RREQ-RREP measurements and the delay experienced by actual data packets. We address this issue by introducing a DUMMY-RREP phase during route discovery. The source saves the RREP packets it receives in a RREP TABLE and then acquires the RREP for a route from this table to send a stream of DUMMY data packets along the path traversed by this RREP. DUMMY packets efficiently imitate real data packets on a particular path owing to the same size, priority and data rate as real data packets. H is the hop count reported by the RREP. The number of packets comprised in every stream is 2H. The destination computes the average delay $D_{avg}$ of all DUMMY packets received, which is sent through a RREP to the source. The source selects this route and sends data packets only when the average delay reported by this RREP is inside the bound requested by the application. The source performs a linear back-off and sends the DUMMY stream on a different route selected from its RREP TABLE when the delay exceeds the required limit

*F .Estimating Data Rate*

In heterogeneous ad hoc networks, throughput through a given route is depending on the minimum data rate of all its links. In a route of links with various data rates, if a high data rate node forwards more traffic to a low data rate node, there is a chance of congestion. This leads to long queuing delays in such routes.

Since congestion significantly reduces the effective bandwidth of a link, the effective link data-rate is given by

$$D_{rate} = D_{Size} / C_{delay} \quad (5)$$

Where $D_{Size}$ is the data size and $C_{delay}$ is the channel delay.

## IV. CONGESTION AWARE ROUTING PROTOCOL (CARP)

CARP is an on-demand routing protocol that aims to create congestion-free routes by making use of information gathered from the MAC layer. CARP employs a combined weight metric in its standard cost function to account for the congestion level.

For establishing multiple disjoint paths, we adapt the idea from the Adhoc On-Demand Multipath Distance Vector Routing (AOMDV) [13]. The multiple paths are computed during the route discovery.

We now calculate the node weight metric $NW$ which assigns a cost to each link in the network. The node weight $NW$ combines the link quality $L_q$, MAC overhead $OH_{MAC}$, effective data rate $D_{rate}$ and the average delay $D_{avg}$, to select maximum throughput paths, avoiding the most congested links.

For an intermediate node $i$ with established transmission with several of its neighbours, the $NW$ for the link from node $i$ to a particular neighboring node is given by

$$NW = (L_q * D_{rate}) \big/ (OH_{MAC} * D_{avg}) \quad (6)$$

*A. Route Request*

Consider the scenario

Let us consider the route

$$S - - N1 - N2 - N3 - D$$

To initiate congestion-aware routing discovery, the source node $S$ sends a RREQ. When the intermediate node $N1$ receives the RREQ packet, it first estimates all the metrics as described in the previous section.

The node $N1$ then calculates its node weight $NW_{N1}$ using (6).

$$RREQ_{N1} \xrightarrow{NW_{N1}} N2$$

$N2$ then calculates its weight $NW_{N2}$ in the same way and adds it to the weight of $N1$. $N2$ then forward the RREQ packet with this added weight.

$$RREQ_{N2} \xrightarrow{NW_{N1}+NW_{N2}} N3$$

Finally the RREQ reaches the destination node $D$ with the sum of node weights

$$RREQ_{N3} \xrightarrow{NW_{N1}+NW_{N2}+NW_{N3}} D$$

*B. Route Reply*

The Destination node $D$ sends the route reply packet RREP along with the total node weight to the immediate upstream node $N3$.

$$RREP_D \xrightarrow{NW_{N1}+NW_{N2}+NW_{N3}} N3$$

Now $N3$ calculates its cost $C$ based on the information from RREP as

$$C_{N3} = (NW_{N1} + NW_{N2} + NW_{N3}) - (NW_{N1} + NW_{N2}) \quad (7)$$

By proceeding in the same way, all the intermediate hosts calculate its cost.

On receiving the RREP from all the routes, the source selects the route with minimum cost value.

## V. SIMULATION RESULTS

*A. Simulation Model and Parameters*

We use NS2 to simulate our proposed protocol in our simulation, the channel capacity of mobile hosts is set to the same value: 2 Mbps. We use the distributed coordination function (DCF) of IEEE 802.11 for wireless LANs as the MAC layer protocol. It has the functionality to notify the network layer about link breakage.

In our simulation, 50 mobile nodes move in a 1500 meter x 300 meter rectangular region for 100 seconds simulation time. We assume each node moves independently with the same average speed. All nodes have the same transmission range of 250 meters. In our simulation, the speed is set as 5m/s. The simulated traffic is Constant Bit Rate (CBR). The pause time of the mobile node is varied as 0,10,20,30 and 40.

Our simulation settings and parameters are summarized in table 1.

TABLE I. SIMULATION SETTINGS

| | |
|---|---|
| No. of Nodes | 50 |
| Area Size | 1500 X 300 |
| Mac | 802.11 |
| Radio Range | 250m |
| Simulation Time | 100 sec |
| Traffic Source | CBR |
| Packet Size | 512 |
| Mobility Model | Random Way Point |
| Speed | 5m/s |
| Pause time | 0,10,20,30 and 40 |

*B. Performance Metrics*

We compare our CARP protocol with the AOMDV [6] protocol**.** We evaluate mainly the performance according to the following metrics, by varying the pause time as 0,10,20,30 and 40.

**Control overhead:** The control overhead is defined as the total number of routing control packets normalized by the total number of received data packets.

**Average end-to-end delay:** The end-to-end-delay is averaged over all surviving data packets from the sources to the destinations.

**Average Packet Delivery Ratio:** It is the ratio of the number of packets received successfully and the total number of packets sent

**Throughput:** It is the number of packets received successfully.

**Drop:** It is the number of packets dropped.

*C. Results*

**A. Based on Pausetime**

In our initial experiment, we vary the pausetime as 0,10,20,30 and 40.

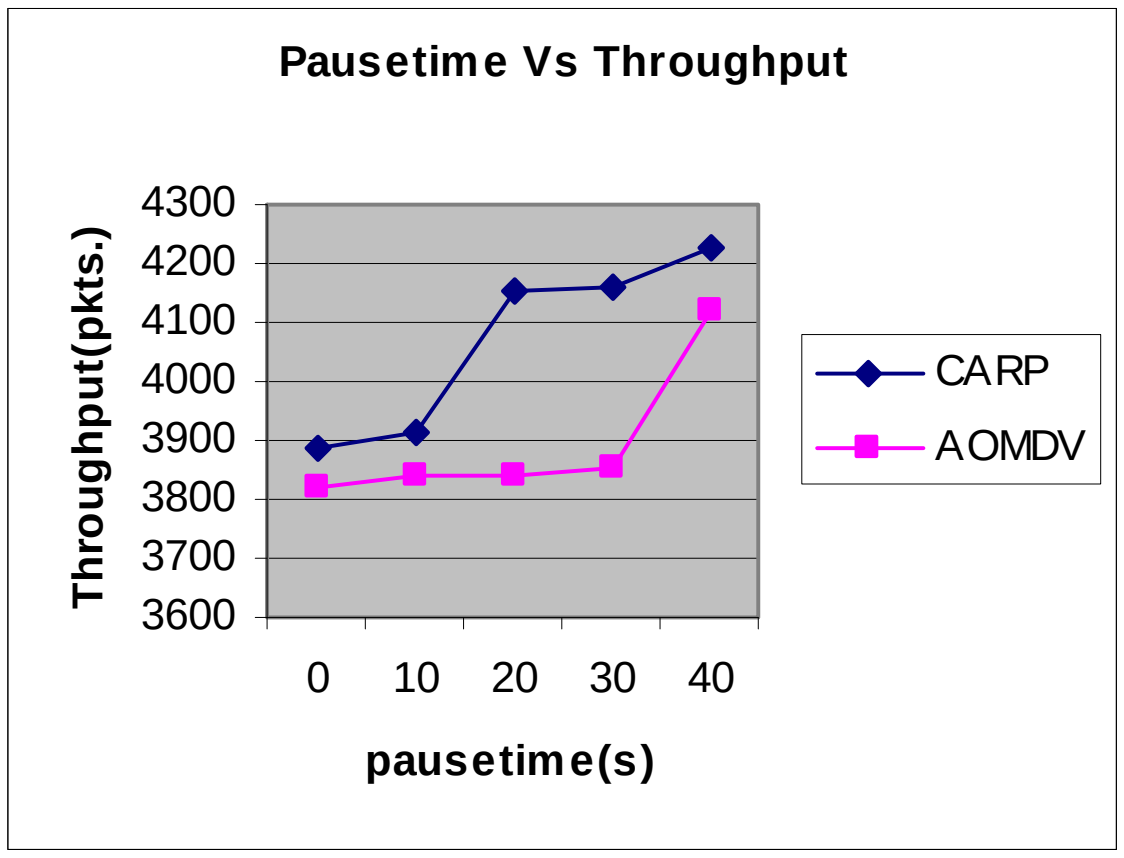

Figure 1. Pausetime Vs Throughput

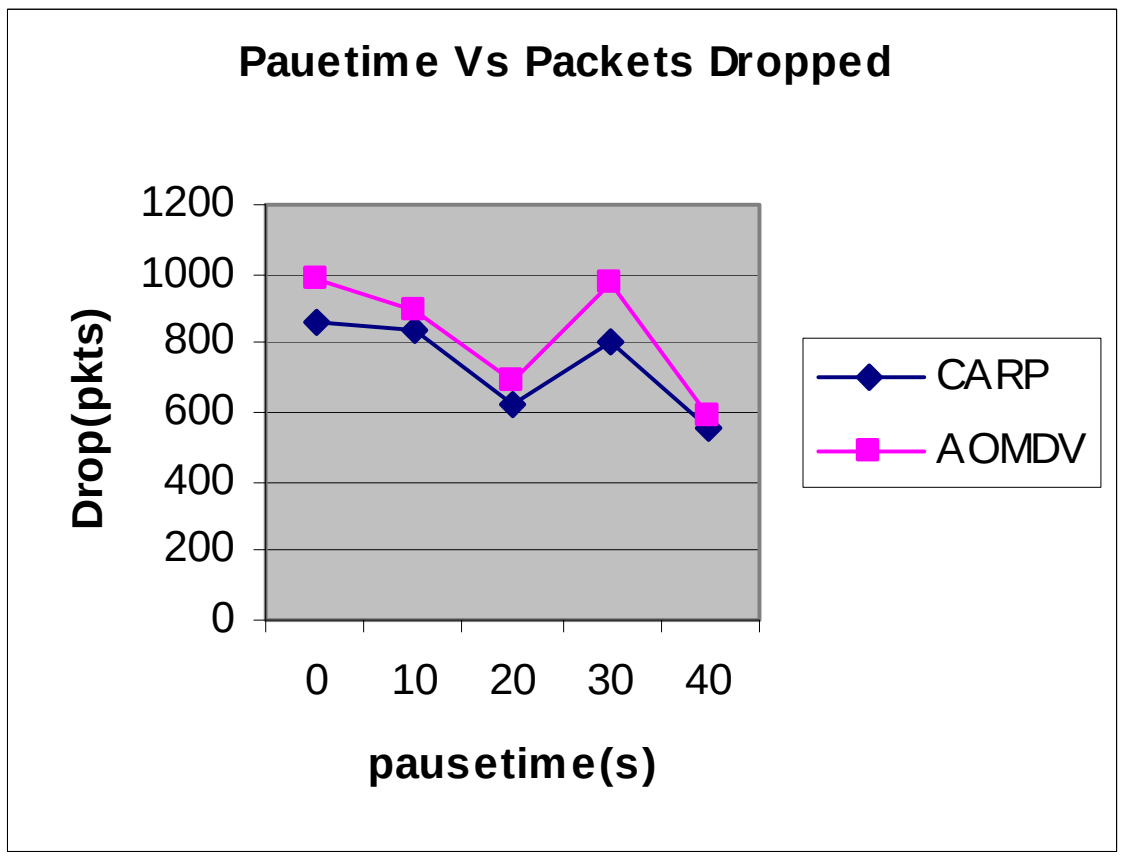

Figure 2. Pausetime Vs Packets Drop

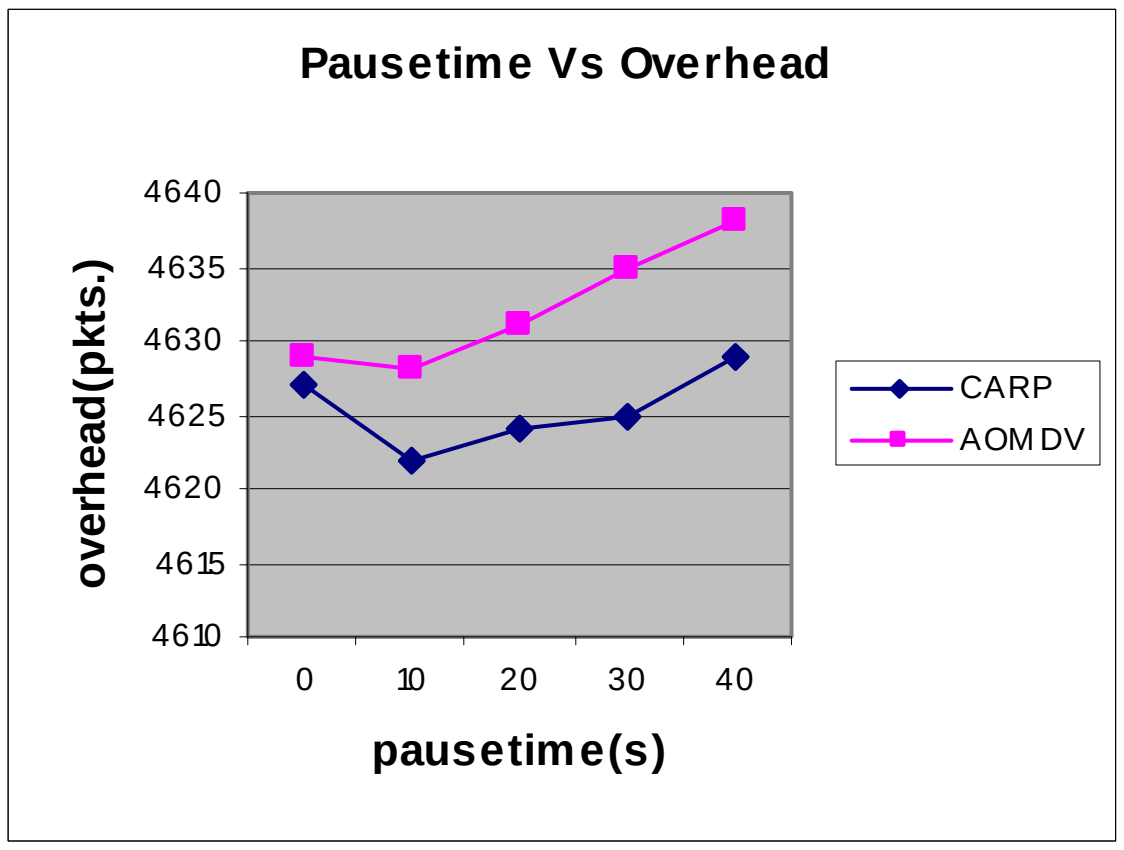

Figure 3. Pausetime Vs Overhead

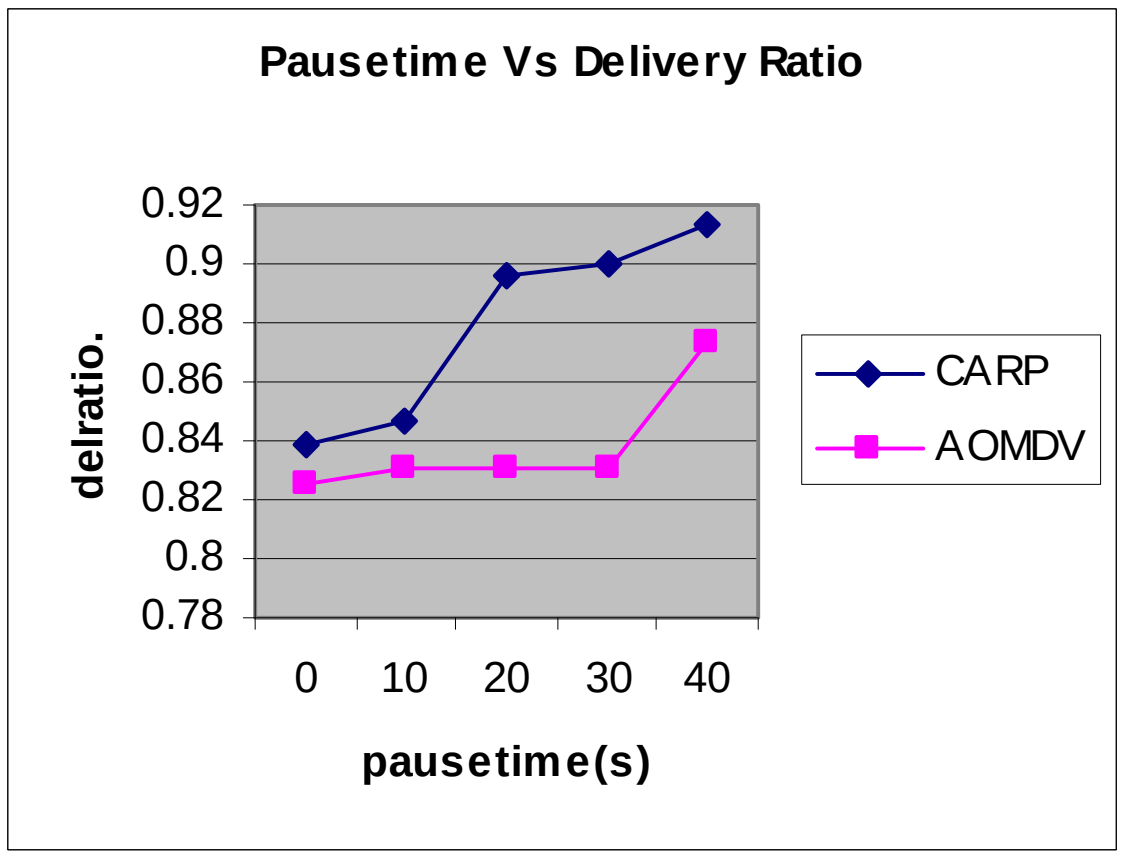

Figure 4. Pausetime Vs Delivery Ratio

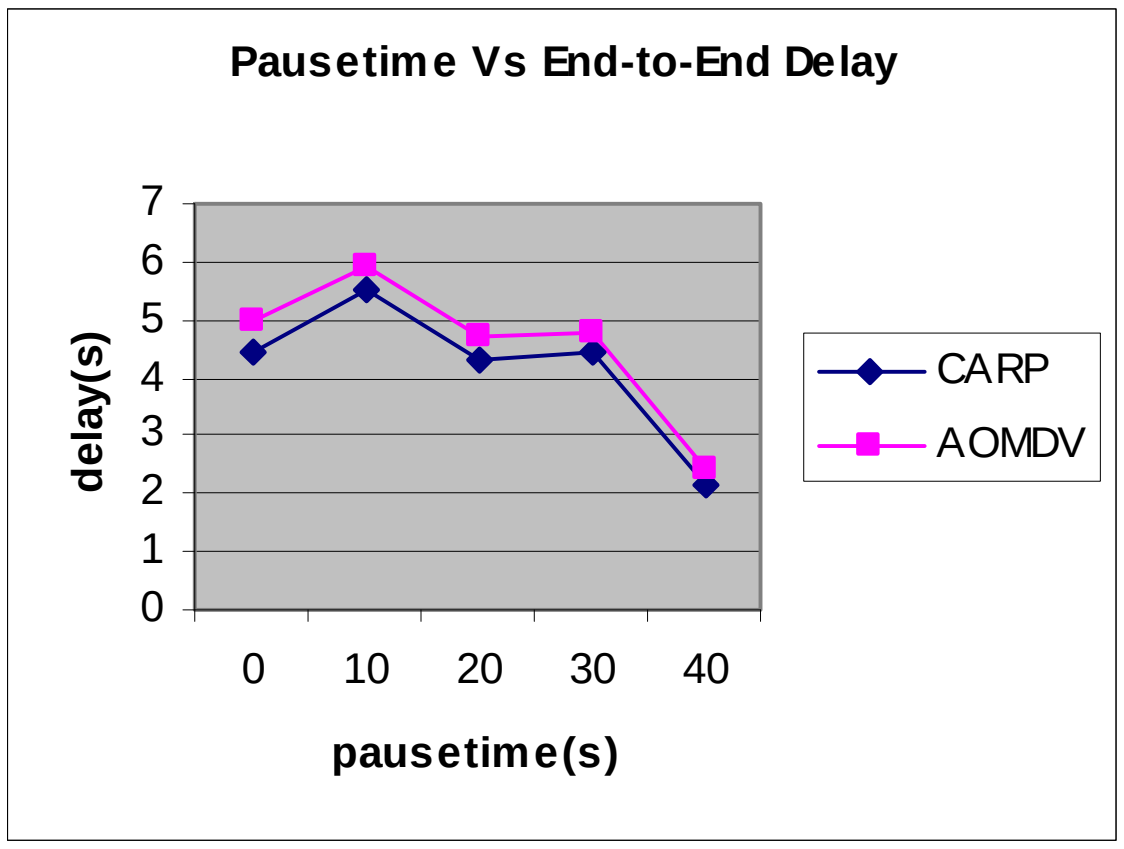

Figure 5. Pausetime Vs Delay

Figure 1 gives the throughput of both the protocols when the pause time is increased. As we can see from the figure, the throughput is more in the case of CARP, than AOMDV.

From Figures 2 and 3, we can ensure that the packets dropped and control overhead is less for CARP when compared to AOMDV.

Figure 4 presents the packet delivery ratio of both the protocols. Since the packet drop is less and the throughput is more, CARP achieves good delivery ratio, compared to AOMDV.

From Figure 5, we can see that the average end-to-end delay of the proposed CARP protocol is less when compared to the AOMDV protocol.

B. Based On Number of Nodes

In the second experiment, we vary the number of nodes as 25, 50, 75 and 100.

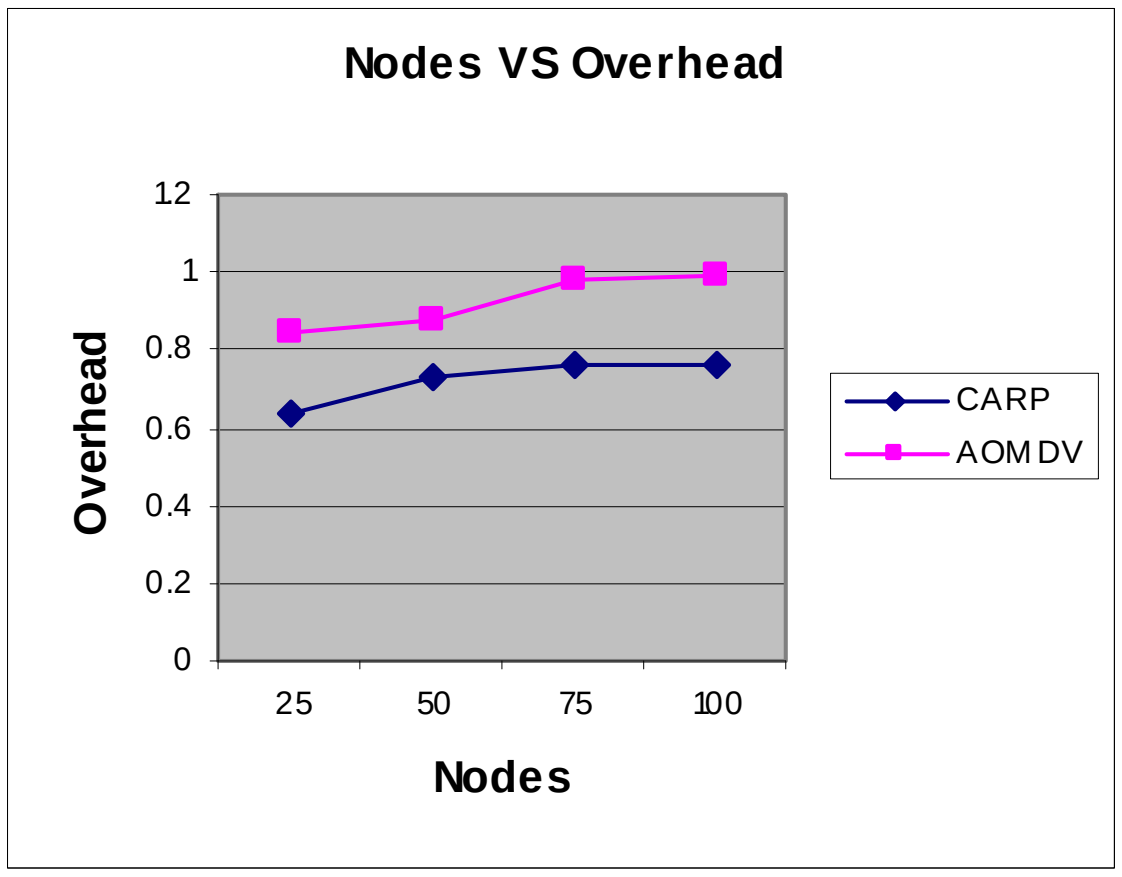

Figure 6. Nodes Vs Overhead

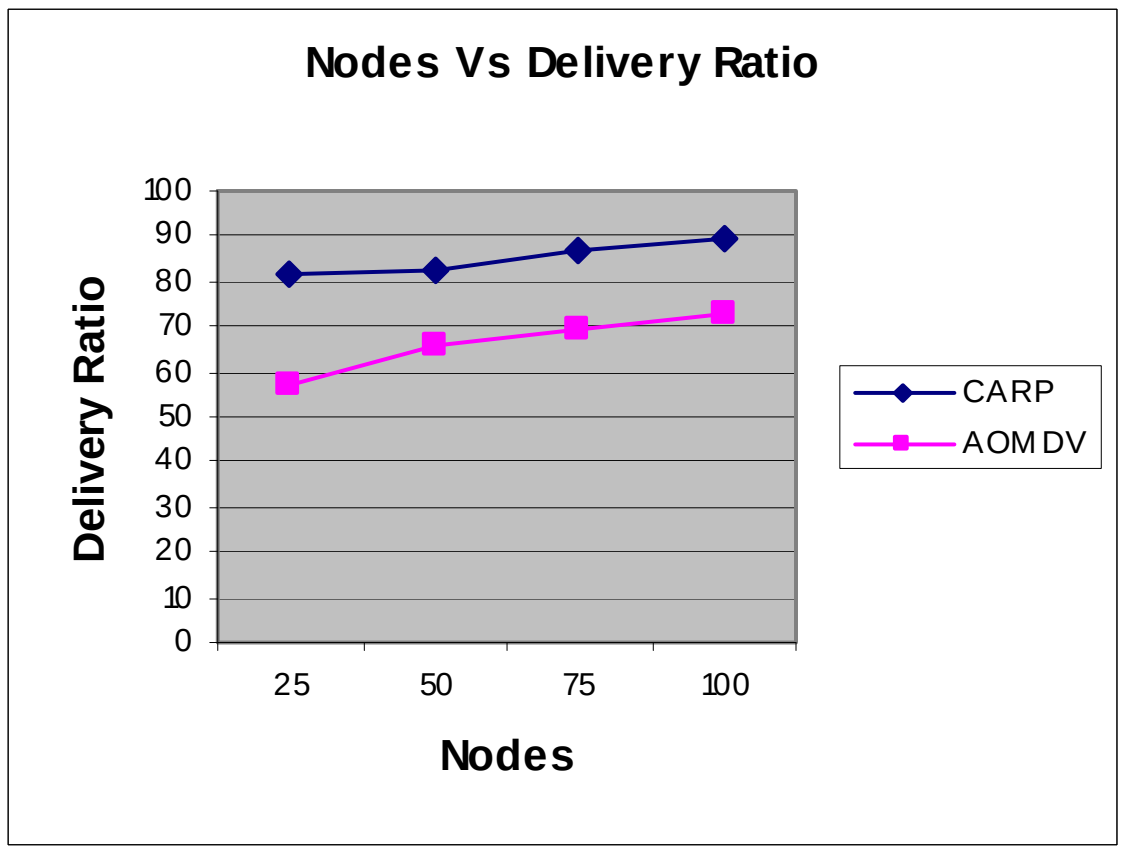

Figure 7. Nodes Vs Delivery Ratio

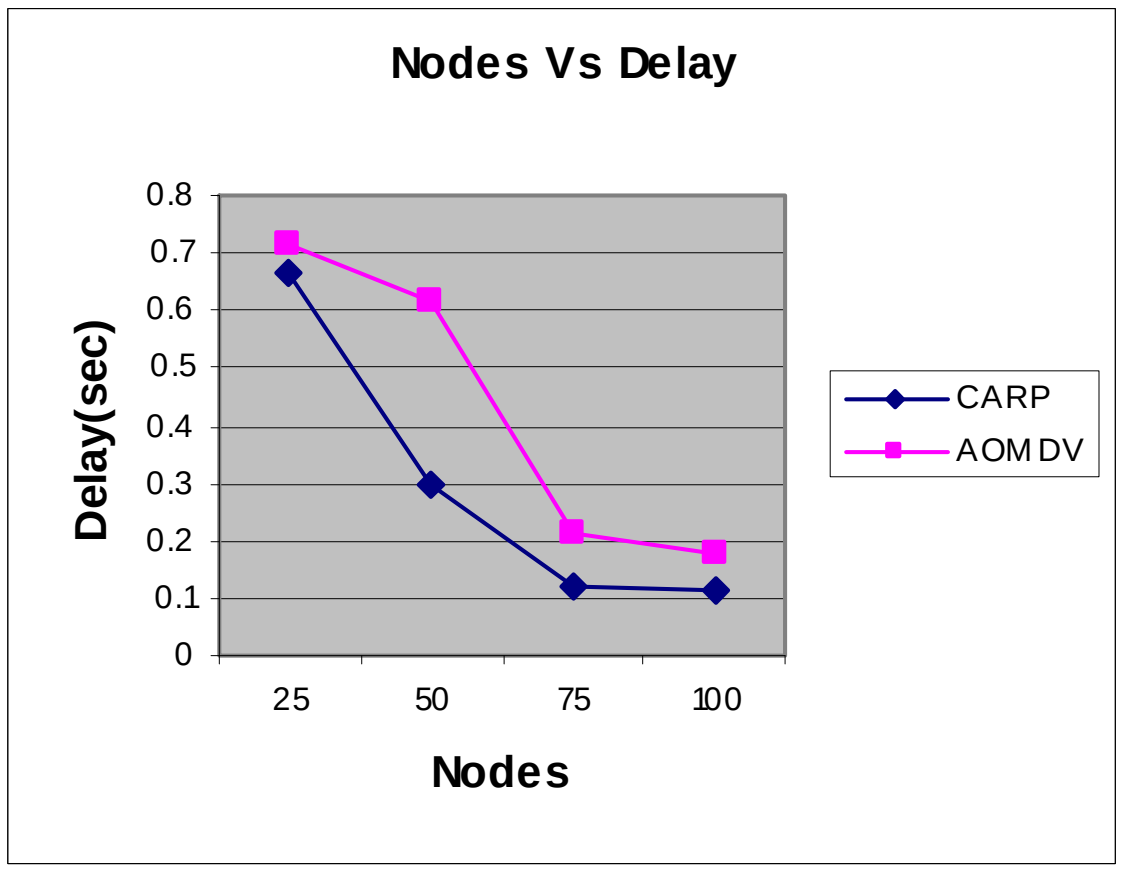

Figure 8. Nodes Vs Delay

From Figures 6, we can ensure that the control overhead is less for CARP when compared to AOMDV.

Figure 7 presents the packet delivery ratio of both the protocols. From the figure we can observe that CARP achieves good delivery ratio, compared to AOMDV.

From Figure 8, we can see that the average end-to-end delay of the proposed CARP protocol is less when compared to the AOMDV protocol.

## VI. CONCLUSION

In heterogeneous mobile ad hoc networks (MANETs) congestion occurs with limited resources and throughput via a given route is depending on the minimum data rate of all its links. In a route of links with various data rates, if a high data rate node forwards more traffic to a low data rate node, there is a chance of congestion, which leads to long queuing delays in such routes. Traditional routing protocols using hop count as a routing metric, do not adapt well to mobile nodes. So there is a need for a congestion aware routing metric which incorporates transmission capability, reliability, and congestion around a link.

In this paper, we have developed a hop-by-hop congestion aware routing protocol which employs a combined weight value as a routing metric, based on the data rate, queuing delay, link quality and MAC overhead. We have used a multipath on demand routing protocol which discovers multiple disjoint routes from a source to destination, as our basis. Among the discovered routes, the route with minimum cost index is selected, which is based on the node weight of all the in-network nodes from the source node to the destination node. By simulation results, we have proved that our proposed routing protocol attains high throughput and packet delivery ratio, by reducing the packet drop and delay.

## AUTHORS PROFILE

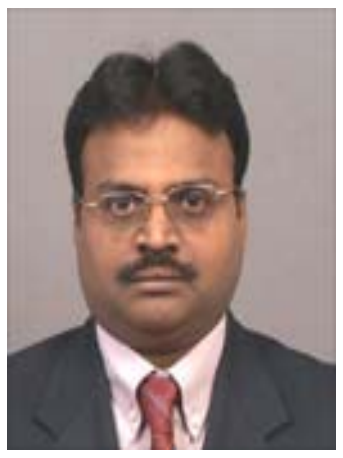

Lt.Dr.S.Santhosh Baboo, aged forty, has around Seventeen years of postgraduate teaching experience in Computer Science, which includes Six years of administrative experience. He is a member, board of studies, in several autonomous colleges, and designs the curriculum of undergraduate and postgraduate programmes. He is a consultant for starting new courses, setting up computer labs, and recruiting lecturers for many colleges. Equipped with a Masters degree in Computer Science and a Doctorate in Computer Science, he is a visiting faculty to IT companies. It is customary to see him at several national/international conferences and training programmes, both as a participant and as a resource person. He has been keenly involved in organizing training programmes for students and faculty members. His good rapport with the IT companies has been instrumental in on/off campus interviews, and has helped the post graduate students to get real time projects. He has also guided many such live projects. Lt.Dr. Santhosh Baboo has authored a commendable number of research papers in international/national Conference/journals and also guides research scholars in Computer Science. Currently he is Senior Lecturer in the Postgraduate and Research department of Computer Science at Dwaraka Doss Goverdhan Doss Vaishnav College (accredited at 'A' grade by NAAC), one of the premier institutions in Chennai..

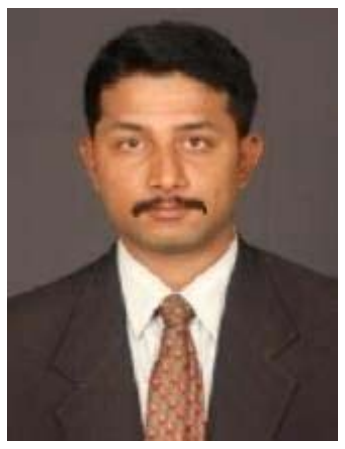

B.Narasimhan, done his Under-Graduation in Annamalai University and Post-Graduation and Master of Philosophy Degrees in School of Computer Science and Engineering, Bharathiar University. He is currently pursuing his Ph.D in Computer Science in Dravidian University, Kuppam, Andhra Pradesh. Also, he is working as a Lecturer, Department of BCA, KG College of Arts and Science. He is having more than one year of research experience and 6 months of teaching experience. His research interest includes Mobile Ad-Hoc Networks and Soft Computing.